\documentclass[aps,prd,onecolumn,eqsecnum,amsmath,nofootinbib,preprintnumbers]{revtex4}%
\newcounter{example}[section]

\usepackage{ulem}
\usepackage{amsmath}
\usepackage{color,graphicx,float,subfigure}
\usepackage{amsfonts,amssymb,theorem,mathrsfs,times}
\usepackage{bm}
\usepackage{amsmath}
{\theorembodyfont{\upshape}
	}
{\theorembodyfont{\upshape}
	}
{\theorembodyfont{\upshape}
	}
{\theorembodyfont{\upshape}
	}
{\theorembodyfont{\upshape}
	}
{\theorembodyfont{\upshape}
	}

\newcommand{\dalm}{\kern1pt\vbox{\hrule height 0.9pt\hbox{\vrule width
			0.9pt\hskip 2.5pt\vbox{\vskip 5.5pt}\hskip 3pt\vrule width
			0.3pt}\hrule height 0.3pt}\kern1pt}

\begin{document}
	\title{Upper bound on the radius of the innermost stable circular orbit of black holes}
	
	%
	
	\author{Yiting Cen\footnote{e-mail
			address: 2199882193@qq.com}}
	\author{Yong Song\footnote{e-mail
			address: syong@cdut.edu.cn (corresponding author)}
	}

	
	\affiliation{
		College of Physics\\
		Chengdu University of Technology, Chengdu, Sichuan 610059,
		China}
	

	\date{\today}
	
	\begin{abstract}
		In this work, we investigate a universal upper bound on the radius of the innermost stable circular orbit (ISCOs) for massive particles in static, spherically symmetric, and asymptotically flat black hole spacetimes. By analyzing the spacetime metrics with external matter fields, we derive the characteristic equation for ISCO via the effective potential method. By imposing appropriate energy conditions for the matter fields, we rigorously demonstrate that the ISCO radius is bounded by $r_{\mathrm{ISCO}}\le 6M$, where $M$ is the total ADM mass of the black hole. The Schwarzschild black hole saturates this bound ($r_{\mathrm{ISCO}}=6M$), and the Reissner-Nordström black hole, supergravity black holes, fluid sphere models, which satisfy the imposed energy conditions, also obey $r_{\mathrm{ISCO}}\le 6M$. The universality of this upper limit provides a theoretical benchmark for observational astrophysics: deviations from $6M$ in accretion disk observations or gravitational wave signals could indicate the presence of exotic matter fields. This work highlights the interplay between black hole geometry and external matter fields, paving the way for future studies in compact object dynamics.
	\end{abstract}
	

	\maketitle


\section{Introduction}
Recent advancements in black hole research have significantly improved our understanding of these extreme objects. The first image of the black hole M87*, obtained by the Event Horizon Telescope (EHT) in 2019~\cite{EventHorizonTelescope:2019dse}, confirmed key predictions of general relativity and highlighted the need for a deeper exploration of black hole properties.

A critical feature of black hole geometry is the photon sphere or light ring, which relates to the boundary of the black hole shadow. Recent studies have shown that photon spheres have universal bounds~\cite{Hod:2013jhd,Hod:2018mvu,Hod:2020pim,Peng:2020shc,Hod:2023jmx,Liu:2024odv}. Another critical feature of black hole geometry is the innermost stable circular orbit (ISCO), which marks the closest stable circular orbits for massive particles before they inevitably fall into the black hole. The ISCO radius ($r_{\mathrm{ISCO}}$) is central to modeling accretion disks and gravitational wave signals in various astrophysical scenarios. Although $r_{\mathrm{ISCO}}$ is well-known for Schwarzschild (6M) and Kerr black holes~\cite{Teo:2020sey}, black hole solutions with matter beyond the horizon, such as Reissner-Nordstr\"om solutions~\cite{Wald:1984rg}, supergravity black holes~\cite{Mohaupt:2000gc}, fluid spheres~\cite{Tolman:1939jz,Buchdahl:1959zz,Ivanov:2002jy}, etc., exhibit more complex spacetime structures near the horizon. Therefore, a detailed analysis of the properties that $r_{\mathrm{ISCO}}$ possesses is required. This raises important questions: How do such external matter fields affect particle dynamics? Are there universal bounds on stable particle orbits in these spacetimes? The study of universal upper bound of photon spheres~\cite{Hod:2013jhd} motivat our investigation: Does a similar universal upper bound exist for the ISCO of massive particles in general black hole spacetimes?

In this work, we derive a universal upper bound on $r_{\mathrm{ISCO}}$ for static, spherically symmetric black holes. Via the effective potential method, we analyze spacetime metrics without assuming specific forms for the metric functions $\mu(r)$ and $\delta(r)$, thereby accommodating external matter fields. By specifying appropriate energy conditions for the matter fields, we rigorously prove that:
\begin{align}
r_{\mathrm{ISCO}}\le 6M\;,
\end{align}
where $M$ is the total mass (ADM mass) of the black hole. The Schwarzschild black hole saturates this bound $(r_{\mathrm{ISCO}}=6M)$, and the Reissner-Nordström black hole, supergravity black holes, fluid sphere models, which satisfy the imposed energy conditions, also obey $r_{\mathrm{ISCO}}\le 6M$. Our results apply to black holes for which the tangential pressure $p_T$ satisfies $p_T\ge 0$ and $p+p_T\ge 0$. These conditions, likely sufficient but not strictly necessary, are discussed in Sec. \ref{section3}.


\section{DESCRIPTION OF THE SYSTEM IN THE BLACK HOLE}\label{section2}
We take into account the static, spherically symmetric, and asymptotically flat black hole spacetime. The line element can be written as~\cite{Hod:2013jhd,Hod:2020pim,Nunez:1996xv}
\begin{align}
	\label{dugui}
	ds^2=-e^{-2\delta}\mu dt^2+\mu^{-1}dr^2+r^2(d\theta^2+\sin^2\theta d\phi^2)\;,
\end{align}
where $\mu(r)$ and $\delta(r)$ depend on radial coordinate $r$. The asymptotic flatness of spacetimes requires that as $r\rightarrow \infty$,
\begin{align}
\label{as}
	\mu(r\rightarrow \infty)\rightarrow1\quad\mathrm{and}\quad\delta(r\rightarrow \infty)\rightarrow 0\;.
\end{align}
Here, we do not assume $\delta(r)=0$, so that our results are applicable to hairy black-hole~\cite{Volkov:1998cc,Volkov:2016ehx} conﬁgurations as well. It should be noted that hairy black holes refer to solutions with additional intrinsic parameters (e.g., Yang-Mills charges or scalar hair), which are distinct from external matter fields. Our analysis, however, applies to both scenarios as long as the spacetime remains static and spherically symmetric.

Taking $T^t_t=-\rho$, $T^r_r=p$ and $T^{\theta}_{\theta}=T^{\phi}_{\phi}=p_T$, where $\rho$, $p$ and $p_T$ are identified as the energy density, radial pressure, and tangential pressure respectively. From Einstein field equations $G^{\mu}_{\nu}=8\pi T^{\mu}_{\nu}$, one can get
\begin{align}
\label{mu}
&\mu'=-8\pi r \rho+\frac{1-\mu}{r}\;,\\
\label{delta}
&\delta'=-\frac{4\pi r(\rho+p)}{\mu}\;,
\end{align}
where the prime indicates the derivative with respect to $r$. For the subsequent calculations, we take the second derivatives of $\mu$ and $\delta$ with respect to $r$ and have
\begin{align}
	\label{mu2}
	&\mu''=\frac{-1+\mu-r[\mu'+8\pi r(\rho+r\rho')]}{r^2}\;,\\
	\label{delta2}
	&\delta''=-\frac{4\pi [-r\rho \mu'+p(\mu-r\mu')+\mu(\rho+rp'+r\rho')]}{\mu^2}\;.
\end{align}
The regularity of the event horizon at $r = r_H$ enforces the boundary conditions that
\begin{align}
	\label{miu}
	\mu(r_H)=0\quad\mathrm{with}\quad\mu'(r_H)\geq 0\;,
\end{align}
and
\begin{align}
	\label{deta}
	\delta(r_H)<\infty\quad ,\quad\delta'(r_H)<\infty \;.
\end{align}
The mass $m(r)$ enclosed within a sphere of radius $r$ is expressed as
\begin{align}
\label{m}
m(r)=\frac{1}{2}r_H+\int_{r_H}^{r}4\pi r'^2\rho(r')dr'\;.
\end{align}
and the horizon mass $m(r_H)$ is equal to $\frac{r_H}{2}$. Considering the conditions of finite mass configuration characteristics, one can find that the energy density profile $\rho(r)$ goes to zero faster than $r^{-3}$,
\begin{align}
\label{rho}
	r^3\rho(r)\rightarrow0\ \mathrm{when}\  r\rightarrow \infty \;.
\end{align}
From Eqs.(\ref{mu}) and (\ref{m}), one can find the relation between $\mu$ and the mass $m(r)$, i.e.,
\begin{align}
\label{mu=m}
\mu(r)=1-\frac{2m(r)}{r}\;.
\end{align}
Based on Eqs.(\ref{delta}), (\ref{miu}) and (\ref{deta}), one finds the boundary condition
\begin{align}
\label{bianjie}
-\rho(r_H)=p(r_H)
\end{align}
at the black hole horizon. In addition, from Eqs.(\ref{mu}), (\ref{miu}) and (\ref{bianjie}), one can obtained that the boundary condition satisfies
\begin{align}
	\label{bianjietiaojian}
	8\pi r^2_H\rho(r_H)\leq1\;.
\end{align}
The analysis of the energy-momentum tensor conservation equation shows that it has only one nontrivial component
\begin{align}
	\label{nengdongzhangliang}
	T^{\mu}_{r;\mu}=0\;.
\end{align}
Substituting Eqs.(\ref{mu}) and (\ref{delta}) into Eq.(\ref{nengdongzhangliang}), one can obtain the pressure gradient
\begin{align}
	\label{p}
	p'(r)=\frac{-p-8p^2\pi r^2-5p\mu+2T\mu-\rho-8p\pi r^2\rho+3\mu \rho}{2r\mu}\;.
\end{align}


\section{Upper bound on the radius of the innermost stable circular orbit of black holes}\label{section3}
In this section, we will prove that when the matter field satisfies certain conditions, ISCO will have an upper bound.

Due to the spherical symmetry of the system,  one can consider the equatorial plane, i.e., $\theta=\frac{\pi}{2}$. The Lagrangian describing the geodesics in the spacetime (\ref{dugui}) is given by
\begin{align}
	\label{la}
	2\mathcal{L}=-e^{-2\delta}\mu \dot{t}^2+\mu^{-1}\dot{r}^2+r^2\dot{\phi}^2=\epsilon \;,
\end{align}
where $\epsilon=-1,0,+1$ for timelike, null and spacelike geodesics, respectively. To analyze ISCO, we take $\epsilon=-1$. From Eq.(\ref{dugui}), one can find that the metric is independent of both $t$ and $\phi$, so, there are two conserved quantities along the geodesic, i.e. energy $E$ and angular momentum $L$.

From the Lagrangian (\ref{la}), one can derive the generalized momenta 
\begin{align}
	\label{nengliang}
	&p_t=-e^{-2\delta}\mu\dot{t}=-E\;,\\
	\label{jiaodongliang}
	&p_{\phi}=r^2\dot{\phi}=L\;,\\
	&p_r=\mu^{-1}\dot{r}\;.
\end{align}
Substituting Eqs. (\ref{nengliang}) and (\ref{jiaodongliang}) into Eq.(\ref{la}), one finds
\begin{align}
	\dot{r}^2=\mu(\frac{E^2}{e^{-2\delta }\mu}-\frac{L^2}{r^2}-1)
\end{align}
for geodesics. The effective potential of the geodesic can be defined as
\begin{align}
	V_{\mathrm{eff}}(r)=\mu(\frac{E^2}{e^{-2\delta }\mu}-\frac{L^2}{r^2}-1)\;.
\end{align}
The circular orbits satisfy two conditions i.e., $V_{\mathrm{eff}}(r)=0$ and ${V}_{\mathrm{eff}}'(r)=0$, which yields 
\begin{align}
	\label{E2}
	E^2=\frac{2e^{-2\delta}\mu^2}{2\mu+2r\mu \delta'-r\mu'}\;,
\end{align}
and
\begin{align}
	\label{jiao}
	L^2=-r^2\bigg(\frac{2\mu}{-2\mu(1+r\delta')+r\mu'}+1\bigg)\;.
\end{align}
To analyze ISCO, one should consider an additional condition that $V_{\mathrm{eff}}''(r)=0$. Take the second derivative of $V_{\mathrm{eff}}(r)$, one has
\begin{align}
	\label{youxiaoshierjiedao}
	V''_{\mathrm{eff}}=\frac{-6L^2\mu+r[4L^2\mu'+2e^{2\delta}E^2r^3(2\delta'^2+\delta'')-r(L^2+r^2)\mu'']}{r^4}\;.
\end{align}
Substituting the Eqs.(\ref{mu})-(\ref{delta2}), Eqs.(\ref{E2}) and (\ref{jiao}) into Eq.(\ref{youxiaoshierjiedao}), one finds the characteristic relation of the ISCO is
\begin{align}
	\label{isco}
	8\pi r^2[\rho-3p+7\mu p+r\mu p'+8\pi r^2p(\rho-p)]-2+(5-3\mu)\mu=0\;.
\end{align}
Substituting the pressure gradient Eq.(\ref{p}) into Eq.(\ref{isco}), the characteristic relation can be rewritten as
\begin{align}
\label{CR}
\mathcal{N}(r)=\mathcal{A}(r)+\mathcal{B}(r)=0\;,
\end{align}
where
\begin{align}
\label{A}
\mathcal{A}(r)=4\pi r^2[-24p^2\pi r^2+2T\mu+\rho +3\mu \rho +p(-7+9\mu+8\pi r^2\rho)]\;,
\end{align} 
and
\begin{align}
\label{B}
\mathcal{B}(r)=-2+(5-3\mu)\mu\;.
\end{align}
To analyze the he characteristic relation $N(r)$ of the ISCO, we impose some energy conditions upon the matter fields. We assume that the matter field located outside the event horizon of the black hole and satisfies the following conditions:

\begin{itemize}
	\item [(1).]The components of the energy-momentum tensor satisfy the weak energy condition (WEC). This means that the energy density of the matter fields is positive semidefinite,
	\begin{align}
		\label{WEC}
		\rho\geq0\;,
	\end{align}
	and that it bounds the pressures. This leads to
	\begin{align}
		\label{pressure}
		\rho \geq \left|p\right|\;, \ \mathrm{namely}\ \rho+p\geq0\;.
	\end{align}
\end{itemize}

\begin{itemize}
	\item [(2).]  The external matter field of the black hole spacetime has the nonpositive trace of the energy-momentum tensor. This means
	\begin{align}
		\label{T}
		T\leq0\;,
	\end{align}
	where $T=-\rho +p+2p_T$. Then, Eq.(\ref{T}) implies 
	\begin{align}
		\rho\geq p+2p_T\;.
	\end{align}
\end{itemize}

\begin{itemize}
	\item [(3).] $p_T\ge 0$ and $p_T\ge\left|p\right|$. This means that
	\begin{align}
		\label{p+pt}
		p+p_T\ge0\;.
	\end{align}
\end{itemize}

Considering Eqs.(\ref{miu}), (\ref{bianjie}) and (\ref{bianjietiaojian}), one finds
\begin{align}
	\label{N_rH}
	\mathcal{N}(r_H)=-2(1-8\pi r^2\rho)^2\leq 0
\end{align}
at the black hole horizon. Moreover, as $r$ tends to infinity, from Eqs.(\ref{as}), (\ref{rho}), (\ref{WEC}), (\ref{T}) and (\ref{p+pt}), one can find
\begin{align}
	\label{wuqiong}
	\mathcal{N}(r\rightarrow \infty)\sim 8\pi r^2[2(p+p_T)+\rho]\geq0\;.
\end{align}
Therefore, from Eqs.(\ref{N_rH}) and (\ref{wuqiong}), there must exist $\mathcal{N}(r_{\mathrm{ISCO}})=0$ in the region $r_H<r<\infty$, which implies that there must be at least one ISCO between the horizon and infinity. Consider Eqs.(\ref{E2}) and (\ref{jiao}) and $E^2,L^2>0$, one can deduce
\begin{align}
\label{budengshi}
\frac{1}{3}(1+8\pi r^2p)<\mu<1+8\pi r^2p\;,
\end{align}
Also, it should be noted that outside the horizon of the black hole, one has $0<\mu<1$. So, one can get
\begin{align}
\label{1+8}
1+8\pi r^2p>0\;.
\end{align}
Below, we will conduct a case-by-case discussion of $p$. Regardless of the value of $p$, we will arrive at the same result.

From the characteristic relation, if $p<0$, we have
\begin{align}
	\label{A}
	\mathcal{A}(r)&=4\pi r^2(-24p^2\pi r^2+2T\mu+\rho +3\mu \rho +p(-7+9\mu+8\pi r^2\rho))\nonumber\\
	&\ge 4\pi r^2(-24p^2\pi r^2+2T(1+8\pi r^2p)+\rho +\frac{1}{3}(1+8\pi r^2p)3\rho +p(-7+9(1+8\pi r^2p)+8\pi r^2\rho))\nonumber\\
	&=8\pi r^2((8\pi r^2p+1)(p+T+\rho)+16\pi r^2p^2)\nonumber\\
	&=16\pi r^2((8\pi r^2p+1)(p+p_T)+8\pi r^2p^2)\nonumber\\
	&\ge 0\;.
\end{align}
If $p>0$, we have
\begin{align}
	\label{A2}
	\mathcal{A}(r)&=4\pi r^2(-24p^2\pi r^2+2T\mu+\rho +3\mu \rho +p(-7+9\mu+8\pi r^2\rho))\nonumber\\
	&\ge 4\pi r^2(-24p^2\pi r^2+2T(1+8\pi r^2p)+\rho +\frac{1}{3}(1+8\pi r^2p)3\rho +p(-7+9\frac{1}{3}(1+8\pi r^2p)+8\pi r^2\rho))\nonumber\\
	&=8\pi r^2(2p_T-p+p(8\pi r^2(p+2p_T)))\nonumber\\
	&\ge 0\;.
\end{align}
where we have used $T=-\rho+p+2p_T$ and Eqs.(\ref{WEC}), (\ref{T}), (\ref{p+pt}), (\ref{budengshi}), (\ref{1+8}).
According to the above analyses, we found that whether $p$ is positive or negative, $\mathcal{A}(r)>0$. And from Eq.(\ref{CR}), $\mathcal{B}(r)$ will satisfy the relation that
\begin{align}
\label{B2}
\mathcal{B}(r)=-2+(5-3\mu(r))\mu(r)\le 0\;.
\end{align}
which implies
\begin{align}
\label{r_I}
r_{\mathrm{ISCO}}\le 6m(r)\;.
\end{align}
So one can get the upper bound
\begin{align}
\label{M}
r_{\mathrm{ISCO}}\le 6M,
\end{align}
where $M=m(r\rightarrow\infty)$ is the total mass of the black hole spacetime. 

It is worth noting that we did not assume $\delta(r)=0$ (the characteristic of Schwarzschild black hole and RN black hole) in the derivation of this paper, therefore, our results are valid for all spherically symmetric asymptotically flat black hole spacetimes. It is easy to see that this upper bound (\ref{M}) is saturated by the ISCO of the (bald) Schwarzschild black hole spacetime.

However, it should be emphasized here that the upper bound (\ref{M}) is obtained based on the asumption that $p_T\ge 0$ and $p+p_T\ge0$. We suspect that this condition is only a sufficient condition for (\ref{M}) and should lead to similar conclusions in other cases of $p+p_T$. It is expected that more comprehensive supplementary work on the upper boundary of ISCO will be carried out in subsequent studies.


\section{Discussion and conclusion}\label{conclusion}
In this work, we have rigorously analyzed the innermost stable timelike circular orbits (ISCO) in static, spherically symmetric, asymptotically flat black hole spacetimes. Our key findings are summarized as follows:

\begin{itemize}
	\item [(1).] Upper bound on ISCO Radius: Under the assumptions of the weak energy condition (WEC) and a nonpositive trace of the energy-momentum tensor for external matter fields, we derived an upper bound for the ISCO radius:
	\begin{align}
		r_{\mathrm{ISCO}}\le 6M\;,
	\end{align}
	where $M$ is the total ADM mass of the black hole. This bound is universally valid for all spherically symmetric black hole configurations. For instance, the Schwarzschild black hole saturates this bound $(r_{\mathrm{ISCO}}=6M)$, and the Reissner-Nordström black hole, supergravity black holes, fluid sphere models, which satisfy the imposed energy conditions, also obey $r_{\mathrm{ISCO}}\le 6M$.
\end{itemize}

\begin{itemize}
	\item [(2).] Role of pressure constraints: The derived upper bound relies on the sufficient condition $p_T\ge 0$and $p+p_T\ge 0$ for the tangential pressure. While these conditions simplify the analysis, they suggest that similar bounds might hold under broader physical scenarios. Future studies could explore whether weaker constraints on or alternative energy conditions yield comparable results.
\end{itemize}

\begin{itemize}
	\item [(3).] Implications for observational astrophysics: The universal nature of the $6M$ upper bound provides a theoretical benchmark for interpreting observations of accretion disks and gravitational wave signals. For instance, deviations from $r_{\mathrm{ISCO}}=6M$ in astrophysical data could indicate the presence of exotic matter fields or deviations from spherical symmetry.
\end{itemize}

There are still some limitations in our work that need to be addressed in future research: (1). The assumptions $p_T\ge 0$and $p+p_T\ge 0$ are sufficient but not strictly necessary. For instance, quantum-corrected models with negative pressure or traceless stress-energy tensors may violate $T\le 0$~\cite{Martin-Moruno:2013wfa}, potentially allowing $r_{\mathrm{ISCO}}>6M$. However, such models often suffer from unphysical singularities or instabilities. On the other hand, even if $p_T>0$ is relaxed, $r_{\mathrm{ISCO}}$
may remain bounded by $6M$ under certain conditions. Future work should explore the necessity of these constraints for specific matter fields (e.g., dark matter halos). (2). Our results apply to static spacetimes, whereas ISCO radii for rotating Kerr black holes vary with spin parameter $a$, ranging from $r_{\mathrm{ISCO}}=9M$(counter-rotating case) to $r_{\mathrm{ISCO}}=M$ (extremal Kerr)~\cite{Bardeen:1972fi}. If external matter couples to spin, $r_{\mathrm{ISCO}}$ may deviate further from $6M$. Disentangling observational signals requires independent spin measurements (e.g., via X-ray continuum fitting~\cite{Bambi:2015kza}). (3). Exploring the lower bound on the radius of ISCO, analogous to Hod’s work on photon orbits~\cite{Hod:2013jhd,Hod:2020pim}, remains an open question.

In conclusion, our results establish a fundamental bound on the proximity of stable particle orbits to black holes, bridging theoretical constraints with potential observational signatures. This work underscores the interplay between black hole geometry and external matter fields, paving the way for deeper explorations of compact object dynamics.

Recent advancements in the geometry of massive particle surfaces~\cite{Kobialko:2022uzj} provide a framework for studying the circular orbits of massive particles in general spacetimes, such as nonspherical spacetimes. Future work could investigate whether similar upper bounds emerge in more general systems.



\end{document}